\documentclass[aps,prl,twocolumn,floatfix,showpacs,a4paper,superscriptaddress]{revtex4-1}
\usepackage{graphicx}
\usepackage{amssymb}
\usepackage{amsmath}
\usepackage{bookmark}
\usepackage{hyperref}

\renewcommand{\phi}{\varphi}

\begin{document}

\title{Discovery of topological metamaterials by symmetry relaxation \\ and  smooth topological indicators}
\author{Cyrill B\"osch}
\affiliation{Institute for Geophysics, ETH Zurich, CH-8092
Zurich, Switzerland}

\author{Tena Dub\v{c}ek}
\affiliation{Institute for Theoretical Physics, ETH Zurich, CH-8093
Zurich, Switzerland}

\author{Frank Schindler}
\affiliation{Department of Physics, University of Zurich, Winterthurerstrasse 190, 8057 Zurich, Switzerland}

\author{Andreas Fichtner}
\affiliation{Institute for Geophysics, ETH Zurich, CH-8092
Zurich, Switzerland}
\date{\today}

\author{Marc Serra-Garcia}
\affiliation{Institute for Geophysics, ETH Zurich, CH-8092
Zurich, Switzerland}

\begin{abstract}
Robustness against small perturbations is a crucial feature of topological properties. This robustness is both a source of theoretical interest and a drive for technological applications, but presents a challenge when looking for new topological systems: Small perturbations cannot be used to identify the global direction of change in the topological indices. Here, we overcome this limitation by breaking the symmetries protecting the topology.  The introduction of symmetry-breaking terms causes the topological indices to become non-quantized variables, which are amenable to efficient design algorithms based on gradient methods. We demonstrate this capability by designing discrete and continuous phononic systems realizing conventional and higher-order topological insulators.
\end{abstract}

\maketitle

The key idea in topological physics has remained largely unchanged since its beginnings in the explanation of the Quantum Hall Effect \cite{Klitzing80} by means of the Chern number \cite{TKNN}: We can associate integer-valued topological invariants to the bulk band properties of a material, and these topological invariants in turn predict the material's boundary physics \cite{PhysRevLett.107.186405}. A crucial property of topological invariants is that they are insensitive to smooth deformations, as long as these respect the protecting symmetries of the topological invariant. This equips topological materials with their characteristic robustness, but presents a challenge from a design point of view, as one cannot systematically discover new topological systems by smoothly-modifying a non-topological model in the direction of growth of the topological index. Proof of this challenge is that discovering new topological models is still as much of an art as a science, as exemplified by the diversity of approaches to the design problem. These include engineering the symmetries of the system \cite{Mousavi2015}, exhaustive search of crystal structure databases \cite{TQC}, identifying geometries that mimic known topological tight-binding models over a range of frequencies \cite{MatlackPerturb, SerraGarciaQuadrupole}, using artificial-intelligence constructs such as neural networks \cite{MachineLearningPilozzi,FlorianML}, or optimizing for proxy quantities such as boundary modes \cite{TopoAcousticBandInversion} or energy transfer \cite{PhononicOptimizationWaveguiding, PhononicOptimizationWaveguidingSR} that, while not topological themselves, are frequently associated with nontrivial topology. 

\begin{figure}[!ht]
\centering
\includegraphics{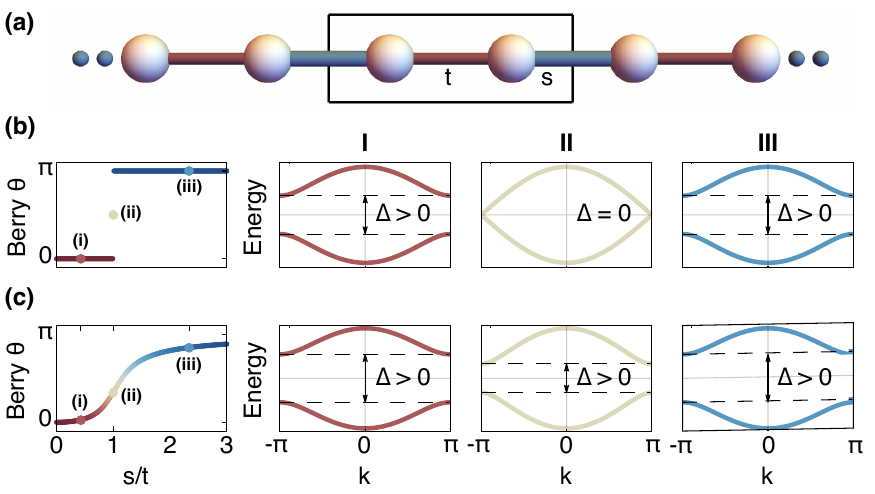}
\caption{Symmetry-relaxation in the SSH model.  (a) Example of an SSH chain. The rectangle highlights the unit cell. The model consists of a dimerized chain with alternating hoppings of strength $t$ and $s$. (b) Berry phase (left) and band structure (right) for an SSH chain with intact inversion symmetry, as the hopping $s$ is varied. (c)  Berry phase (left) and band structure (right) for an SSH chain with relaxed inversion symmetry, as the hopping $s$ is varied. }
\label{fig:sshfig}
\end{figure}

In this paper, we approach the realization of topological phases by breaking the protecting symmetries and therefore challenging the discreteness of topological invariants.  The approach can be understood in the SSH \cite{SSH} model (Fig. \ref{fig:sshfig}a), which consists of a 1-dimensional chain with dimerized hoppings of strength $s$ and $t$. Such a system is described by the Bloch Hamiltonian 

\begin{eqnarray} \label{eq:ssheq}
H=\left( \begin{matrix} 
0 & t + se^{ik}\\
 t + se^{-ik} & 0 
\end{matrix}\right).
\end{eqnarray} When the hoppings follow the inequality $t < s$, the system will be in a topological phase, and finite samples will present boundary-localized states in the gap.  When $t > s$, the system will be in a trivial phase, and finite samples will not present in-gap states at the boundary. This topology can be characterized by a Berry phase invariant of the form

\begin{eqnarray} \label{eq:geophase}
\theta=i\oint{u^*(k) \frac{d}{dk}u(k)dk},
\end{eqnarray} Where $u(k)$ is the eigenfunction of Eq. (\ref{eq:ssheq}) with energy below the gap.

Smoothly deforming the system by altering the couplings $s$ or $t$  (Fig. \ref{fig:sshfig}b) does not affect the topological invariant in Eq. (\ref{eq:ssheq}), unless the system goes through the configuration where $s=t$. In this configuration, the lattice is not dimerized, the gap closes and the topological invariant is not defined.  This picture is, however, only true when the symmetry that protects the topological invariant is preserved -- which in this case corresponds to inversion symmetry. If we eliminate the inversion symmetry, by adding a local potential,

\begin{eqnarray} \label{eq:sshasym}
H_{\mathrm{relaxed}}=\left( \begin{matrix} 
\epsilon & t + se^{ik}\\
 t + se^{-ik} & -\epsilon 
\end{matrix}\right),
\end{eqnarray}we observe two effects (Fig. \ref{fig:sshfig}c): First, the Berry phase changes smoothly when we vary the hoppings $s$ and $t$, and second, the gap $\Delta$ does not close in the configuration where $s = t$, meaning that Eq. (\ref{eq:ssheq}) can be evaluated for all hoppings. While this quantity can no longer be interpreted as a topological invariant, we will show that it can be used in the search process to discover novel topological systems. We refer to these quantities as \emph{smooth topological indicators}.

We will now demonstrate this method by designing a continuous phononic system with non-trivial band gaps, and demonstrate the existence of boundary modes. Classical systems presenting topological wave phenomena  \cite{Wang2009, Hafezi2013, Khanikaev2013,SusstrunkPendulaScience, SimonSchusterCircuit, GlazmanCircuit} are an established platform for the demonstration of novel condensed matter physics. Examples include the demonstrations of fragile topological phases \cite{periFragile}, higher-order topological insulators \cite{BahlMicrowaveQuadrupole, SerraGarciaQuadrupole}, and Weyl semi-metal effects such as axial fields \cite{peri2019axialweyl} and surface physics \cite{AcousticWP2}, which were all first observed in classical models. 
\begin{figure}[!ht]
\centering
\includegraphics{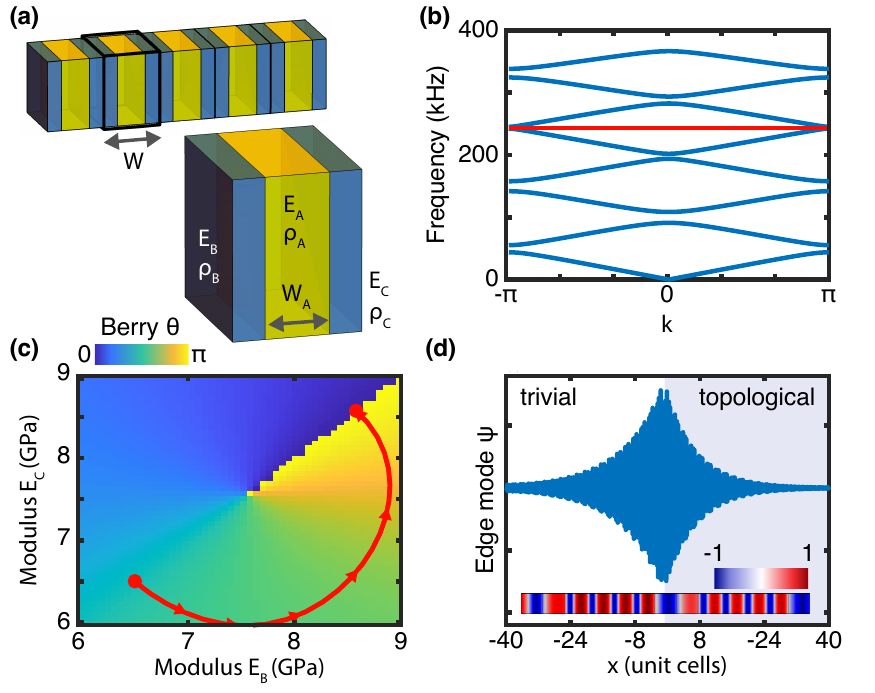}
\caption{Design of a phononic topological insulator using smooth topological indicators.  (a) Material design made by repeating a 3-material unit cell.  (b) Dispersion relation for the system in a. The red region denotes the gap selected for optimization. (c)  Multiband Berry phase as a function of the unit cell parameters $E_b$ and $E_c$. The red line shows a gradient-descending optimization trajectory starting from a symmetric, trivial phase and ending in a symmetric, topological configuration. (d) Finite Element Method simulation (COMSOL\textregistered  Multiphysics) of a localized mode at the boundary between a trivial and a topological material \cite{SupMat}, corresponding respectively to the start-point and endpoint in c. The domain wall is placed at $x=0$. The inset shows the localized mode at the two unit cells adjacent to the domain wall. The moduli $E_A$, $E_B$ and $E_C$ corresponding to the trivial phase have been scaled by a proportionality constant to ensure overlapping band gaps. Except when otherwise indicated, the system parameters are $\rho=2704\,kg/m^3$, $E_A=2\,GPa$, $E_B = E_C = 6.5\,GPa$, $W=10\,mm$ and $W_A=0.6759\,W$.}
\label{fig:topoContinuous}
\end{figure}
The continuous phononic metamaterial considered here is a one-dimensional slab of elastic material, satisfying the elastic wave equation,

\begin{eqnarray} \label{eq:wavequation}
\rho\left(x\right) \partial_{tt}\psi\left(x,t\right) + \partial_{x}\left[E\left(x\right)\partial_{x}\psi\left(x,t\right)\right] = 0,
\end{eqnarray} where $\psi$ is the displacement along the longitudinal direction ($x$) of the slab, $\rho$ is the density and $E$ is the modulus of elasticity. The metamaterial is made by periodically repeating a unit cell consisting of a three-material sandwich \cite{EM1DTopo, elasticString} with constant density $\rho$ and moduli $E_A$, $E_B$ and $E_C$, respectively (Fig. \ref{fig:topoContinuous}a). This system has multiple band gaps (Fig. \ref{fig:topoContinuous}b) We will consider a Berry-phase like topological invariant, generalized to multi-band systems \cite{vanderbilt}, defined by:

\begin{eqnarray} \label{eq:diffdtheta}
\theta = -i\,\log \det  \mathcal{U} ,
\end{eqnarray} where $\mathcal{U}_{ij} = \langle u_i(-\pi) | u_j(\pi) \rangle$ is obtained by parallel-transport of the Bloch wavefunctions $u_i(k)$ from $-\pi$ to $\pi$.

Figure \ref{fig:topoContinuous}c shows the multi-band Berry phase associated to the fifth band gap, for different values of the parameters $E_B$ and $E_C$. Along the line where the $E_B = E_C$, inversion symmetry quantizes the Berry phase to the values of $0$ or $\pm \pi$. However, allowing symmetry violations results in a path through parameter space where the Berry phase transitions smoothly between trivial and topological values, without going through a gap closing. Using a gradient-ascent algorithm, it is therefore possible to identify topological configurations of the phononic system in a direct manner. We demonstrate this ability by identifying parameter values for $E_B$ and $E_C$ resulting in topological and trivial configurations (Fig. \ref{fig:topoContinuous}c) and observing the presence of localized interface modes at the selected band gap in a Finite Element Method simulation of the designed phononic material (Fig. \ref{fig:topoContinuous}d). Two aspects of this result deserve special mention: By having selected realistic initial conditions, we obtain a system geometry and material properties that can be easily realized experimentally. Second, the system ends up naturally in a symmetry-respecting configuration. Symmetry in the final configuration cannot be always guaranteed to appear automatically, but can be restored by penalizing symmetry-breaking terms towards the end of the optimization.

\begin{figure}[!ht]
\centering
\includegraphics{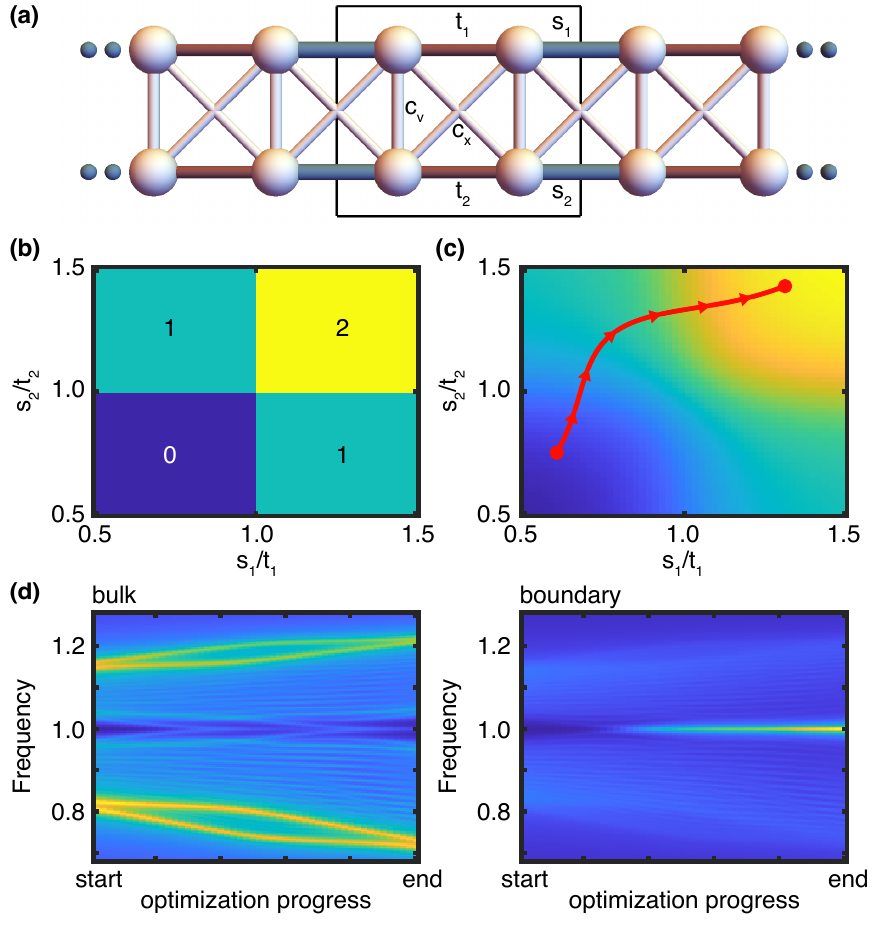}
\caption{Relaxation of topological indices based on symmetry eigenvalues at TRIMs. (a) Double SSH chain with couplings $t_1$,  $t_2$, $s_1$, $s_2$, $c_v$ and $c_x$. (b) Topological index in the presence of inversion symmetry. (c) Topological index after the symmetry has been lifted by perturbations $\epsilon_i\neq0$. The red line represents a gradient-descending optimization trajectory. (d) Bulk and boundary responsivity $\Psi_i^2\left(\omega\right)$ of a finite sample, as the hopping strengths are being optimized from a trivial to a topological configuration following the red line in c.}
\label{fig:topoSymmetry}
\end{figure}

Topological insulators with time-reversal and inversion symmetries can also be characterized through the eigenvalues of the parity operator evaluated at the Time-Reversal Invariant Momenta (TRIM) points in reciprocal space \cite{FuInversionTI}. This type of invariant can also be smoothed out by symmetry relaxation, as we demonstrate in the double-SSH model. The double-SSH model consists of two parallel coupled SSH chains (Fig. \ref{fig:topoSymmetry}a), and cannot be characterized with a conventional Berry phase, which is zero for both topological and trivial configurations. The system is described by the Hamiltonian

\begin{eqnarray} \label{eq:doubSSHEQ}
\begin{split}
& H_{D}= \\ &\left( \begin{matrix} 
V+\epsilon_1 & t_1 + s_1e^{ik} & c_v+c_ve^{ik}  & c_x+c_xe^{ik} \\
 t_1 + s_1e^{-ik} & V-\epsilon_1 & c_x+c_xe^{ik} & c_v+c_ve^{ik} \\
 c_v+c_ve^{-ik} & c_x+c_xe^{-ik}  & V+\epsilon_2 & t_2 + s_2e^{ik} \\
  c_x+c_xe^{-ik} &  c_v+c_ve^{-ik}  & t_2 + s_2e^{-ik} & V-\epsilon_2
\end{matrix}\right),
\end{split}
\end{eqnarray} where $V$ is a local potential ensuring postive-definiteness of the classical model, $t_1$, $t_2$, $s_1$ and $s_2$ are the real-valued hopping parameters for the two chains, $c_x$ and $c_v$ are the real-valued chain coupling strengths and $\epsilon_1$ and $\epsilon_2$ are real-valued symmetry-breaking terms introduced to obtain a smooth topological indicator.

The topological invariant is defined as $\nu =  n_-(0) - n_-(\pi)$  \cite{WilsonLoopInversion} where $n_-\left(k\right)$ is the number of Bloch eigenfunctions with $-1$ parity eigenvalue, evaluated at the point $k$. For inversion-symmetric systems at TRIMs, the parity eigenfunctions can only take values $+1$ or $-1$, and the invariant $\nu$ reduces to the form

\begin{eqnarray} \label{eq:invariantHOTI}
\begin{split}
\nu = \frac{1}{2}  \mathrm{Tr} \left[ U^\dagger (0)\left(\hat{I}-\hat{P}\right) U (0) \right] \\-   \frac{1}{2}\mathrm{Tr}\left[ U^\dagger (\pi)\left(\hat{I}-\hat{P}\right) U (\pi) \right],
\end{split}
\end{eqnarray}  where $\hat{I}$ is the identity operator, $\hat{P}$ is a parity transformation and $U (k)$ is a matrix whose columns are the Bloch eigenfunctions below the band gap of interest, evaluated at the momentum point $k$. The function in Eq. (\ref{eq:invariantHOTI}) is gauge-invariant and becomes quantized when the inversion symmetry ($\epsilon_1 = \epsilon_2 = 0$) is respected (Fig. \ref{fig:topoSymmetry}b).

If we lift the inversion symmetry by setting $\epsilon_1 \neq 0$ and  $\epsilon_2 \neq 0$, the value of Eq. (\ref{eq:invariantHOTI}) becomes smooth (Fig. \ref{fig:topoSymmetry}c). Under these conditions, $\nu$ cannot be interpreted as a topological quantity. However, since $\nu$ is smooth and equals the topological invariant when symmetry is respected, it can be used to guide a search algorithm towards a topological configuration.  This process is illustrated in Fig. \ref{fig:topoSymmetry}c and Fig. \ref{fig:topoSymmetry}d for a 20-site, classical implementation of the tight-binding model in Fig. \ref{fig:topoSymmetry}a, where each site is implemented as a mass $m$ with local damping $b$. As the gradient-ascent algorithm increases the smooth topological indicator, the frequency response of the finite, symmetry-preserving system nucleates boundary modes. Two distinct transitions can be observed in Fig. \ref{fig:topoSymmetry}d, corresponding to the points where $s_2 = t_2$ and $s_1 = t_1$.

Design by symmetry relaxation can also be applied to Higher-Order Topological Insulators (HOTIs), systems where the non-trivial bulk topology gives rise to boundary modes that are more than one dimension below that of the bulk \cite{SchindlerHOTI, BenalcazarMultipole}. Two-dimensional HOTIs can be protected by mirror symmetries $\mathcal{M}_x$ and  $\mathcal{M}_y$. For such systems, we introduce a topological characterization \cite{SupMat} based on four mirror-graded invariants \cite{WilsonLoopInversion, BerryPhaseCrystaline, FractionalCorner} of the form $\nu_\pm^x = \left| \theta_\pm(k_x=0) -  \theta_\pm(k_x=\pi)\right|$ and $\nu_\pm^y = \left| \theta_\pm(k_y=0) -  \theta_\pm(k_y=\pi)\right|$, with the topological phase characterized by $\nu_\pm^x=\nu_\pm^y=\pi$. The quantities  $\theta_\pm$ are multiband Berry phases calculated on a sub-set of the bands below the gap of interest, selected according to the band symmetry characteristics, with $\theta_+$ ($\theta_-$) being the multiband Berry phase calculated using the bands with positive (negative) eigenvalue according to the classifying symmetry. When calculating the invariants $\nu_\pm^x$ ($\nu_\pm^y$), the classifying symmetry will be $\mathcal{M}_x$ ($\mathcal{M}_y$). The multiband Berry phases that appear in the expression of $\nu_\pm^x$ ($\nu_\pm^y$) are quantized by a different mirror symmetry, $\mathcal{M}_y$ ($\mathcal{M}_x$), as the one used for classification. Therefore, by lifting one of the symmetries while respecting the other, a pair of invariants becomes a well-defined smooth topological indicator.

\begin{figure}[!ht]
\centering
\includegraphics{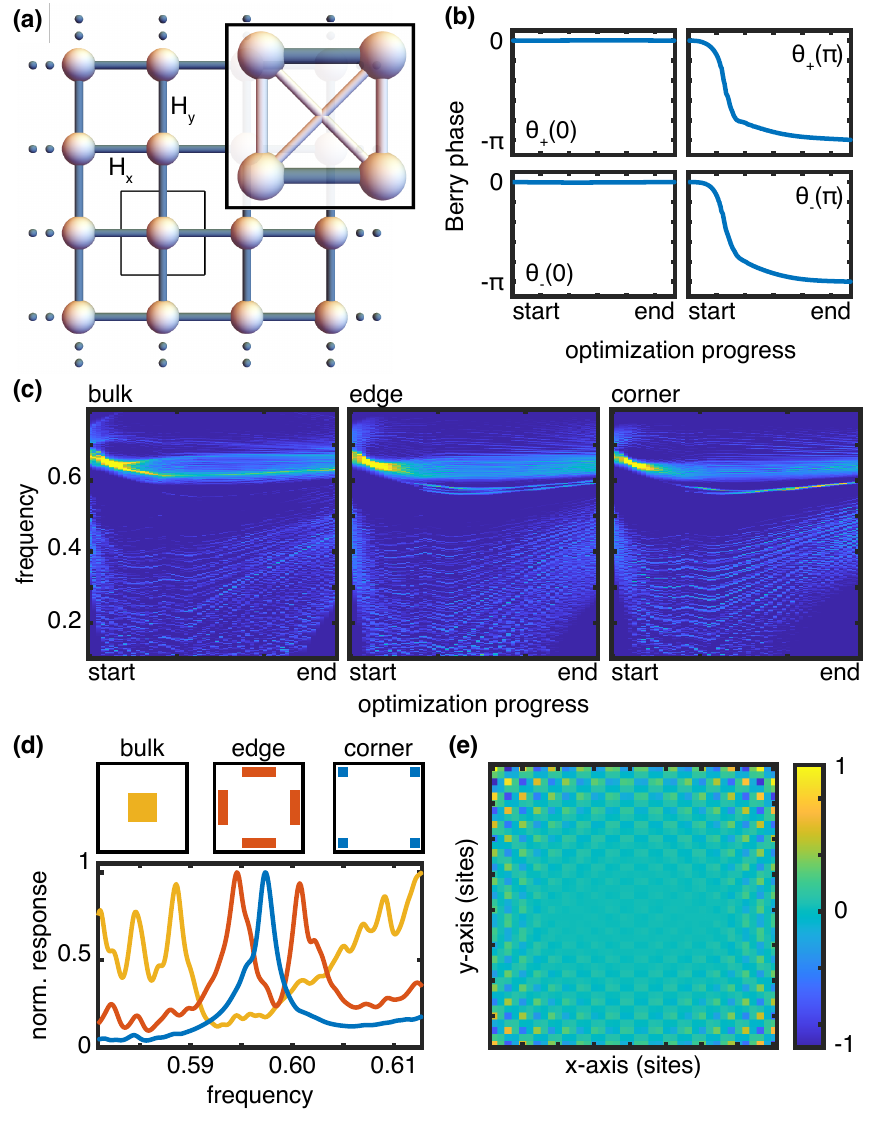}
\caption{Symmetry relaxation in higher-order topological insulators (a) Square lattice with four sites per unit cell (inset). Each unit cell interacts only with nearest-neighbor unit cells.  (b) Evolution of the topological invariants $\theta_\pm(k)$ during the design process. (c)  Evolution of the frequency spectrum for a finite, classical system consisting of 20x20 unit cells, evaluated at the bulk, edges and corners.  (d) Bulk, edge and corner responses of the finite classical sample in c, at the end of the design process. (e) Example of a corner-localized eigenmode of the finite system in c. In panels c-d, a damping of $b=0.05-0.02$ has been used.}
\label{fig:topoHighOrder}
\end{figure}

Smoothed mirror-graded invariants can be used to design HOTIs in a straightforward way. We will demonstrate this in a square lattice with four sites per unit cell (Fig. \ref{fig:topoHighOrder}a). Interactions are restricted to nearest-neighbor unit cells, but not to nearest-neighbor sites. The system is described by a Hamiltonian of the form

\begin{eqnarray} \label{eq:hamiltonianHOTI}
H\left(k_x, k_y \right) = H_0 + e^{ik_x}H_x + e^{ik_y}H_y + h.c.,
\end{eqnarray}  with $H_0$ representing the potentials and interactions inside the unit cell, and $H_x$, $H_y$ the hoppings between unit cells in the $x$ and $y$ direction, respectively. It should be noted that the model does not require any prior knowledge of higher-order topological models. It must only respect the appropriate symmetries and provide a large enough search space so that a topological solution can be identified.

The optimization process for the invariants $\nu_\pm^x$ starts from a highly symmetric configuration that respects  $\mathcal{M}_x$,  $\mathcal{M}_y$ and $\mathcal{C}_4$. During optimization, the classifying symmetry $\mathcal{M}_x$ is respected, while the quantizing symmetry $\mathcal{M}_y$ is allowed to relax. The matrices of the system are optimized following a gradient-ascent algorithm, with two additional requirements: First, the direction of change is required to be orthogonal to the gradient of the band gap size --meaning that changes in the system parameters are not allowed to alter the gap size. This is done to prevent the system from undergoing a gap closing during optimization, which could result in an exchange of classifying symmetry eigenvalues. Second, as the optimization progresses, violations of the $\mathcal{M}_y$ symmetry are increasingly penalized, to compensate for the fact that the system does not naturally converge to a symmetric configuration as it was the case in previous examples. As shown in the Supplemental Material \cite{SupMat}, the presence of the $\mathcal{M}_x$ and $\mathcal{M}_y$ symmetries causes the topological invariants along $y$ to be equal to those along $x$, and therefore the system ends up in a fully topological configuration. In this example, the optimization process can get trapped in local-minima. Here, we escape them by re-starting the optimization from a different random configuration (see supplementary materials for source code implementing the exact algorithms and parameters \cite{SupMat}), but for more complex metamaterials, advanced stochastic algorithms such as Hamiltonian Monte Carlo will be more performant \cite{LarsHMC}.

The optimization process is illustrated in Fig. \ref{fig:topoHighOrder}b-c. The four multiband Berry phases converge to quantized values of $0$ or $\pi$ (Fig. \ref{fig:topoHighOrder}b) while the gap remains open (Fig. \ref{fig:topoHighOrder}c), although not fully constant, as one would expect from the imposed orthogonality condition. We attribute this to inaccuracies in the gap gradient estimation due to degeneracies and finite differentiation. As the optimization progresses, edge and corner modes nucleate in the gap (Fig. \ref{fig:topoHighOrder}c). The system ends up presenting the hallmarks of higher-order topology \cite{SerraGarciaQuadrupole}, namely gapped edge states in the the bulk band gap, and corner states in the edge gap (Fig. \ref{fig:topoHighOrder}d) The eigenfunction corresponding to one of such eigenstates is presented in Fig. \ref{fig:topoHighOrder}e.

In conclusion, we have demonstrated that it is possible to derive smooth topological indicators by relaxing a topological system's protecting symmetries, and that these indicators can be used to guide a search process towards topological configurations. The approach is very generic, as shown by the fact that it was successfully applied to discrete and continuous systems, indices based on geometric phases and symmetry eigenvalues, as well as on conventional and higher-order topological insulators. The approach should provide a novel route towards the realization of topological phases in diverse platforms such as photonic and phononic metamaterials or ultracold atoms. In fact, we expect the approach to be applicable to all symmetry-protected topological systems, including crystalline topological insulators \cite{LiangFuTopoCrystalline} or those protected by chiral, time-reversal or parity symmetries \cite{tenfoldway}.

\section{Acknowledgements}

The authors would like to thank Valerio Peri, Eli\v{s}ka Greplova and Sebastian Huber for helpful discussions.

This work was supported by the European Research Council (ERC) under the European Union's Horizon 2020 Research and Innovation Programme (Grant Agreements No. 694407 and 714069). T.D. acknowledges funding from the ETH Zurich Postdoctoral Fellowship program. F.S. was supported by the Swiss National Science Foundation (Grant No. 200021\_169061).

\bibliographystyle{phd-url}
\bibliography{ref}

\end{document}